\documentclass[sigconf,natbib=false]{acmart}

\usepackage{multirow}
\usepackage{longtable}
\usepackage{CJKutf8}
\usepackage{tcolorbox}
\usepackage{enumitem}
\usepackage{subfigure}
\AtBeginDocument{%
  }

\setcopyright{acmlicensed}
\copyrightyear{2018}
\acmYear{2018}
\acmDOI{XXXXXXX.XXXXXXX}

\acmConference[ASE 2024]{39th International Conference on Automated Software Engineering}{Oct 2024}{California, United States}
\acmISBN{978-1-4503-XXXX-X/18/06}

\begin{document}

\title{Coding-PTMs: How to Find Optimal Code Pre-trained Models for Code Embedding in Vulnerability Detection?}


\author{Yu Zhao}
\affiliation{%
  \institution{School of Computer Science and Technology, Nanjing University of Aeronautics and Astronautics}
  \city{Nanjing}
  \country{China}
}
\email{zhao_yu@nuaa.edu.cn}

\author{Lina Gong}
\authornote{Corresponding author.}
\affiliation{%
  \institution{School of Computer Science and Technology, Nanjing University of Aeronautics and Astronautics}
  \city{Nanjing}
  \country{China}
}
\email{gonglina@nuaa.edu.cn}

\author{Zhiqiu Huang}
\affiliation{%
  \institution{School of Computer Science and Technology, Nanjing University of Aeronautics and Astronautics}
  \city{Nanjing}
  \country{China}
}
\email{zqhuang@nuaa.edu.cn}

\author{Yongwei Wang}
\affiliation{%
  \institution{Shanghai Institute for Advanced Study and College of Computer Science, Zhejiang University}
  \city{Shanghai}
  \country{China}
}
\email{yongweiw@outlook.com}

\author{Mingqiang Wei}
\affiliation{%
  \institution{School of Computer Science and Technology, Nanjing University of Aeronautics and Astronautics}
  \city{Nanjing}
  \country{China}
}
\email{mingqiang.wei@gmail.com}

\author{Fei Wu}
\affiliation{%
  \institution{College of Computer Science and Technology, Zhejiang University}
  \city{Hangzhou}
  \country{China}
}
\email{wufei@zju.edu.cn}

\renewcommand{\shortauthors}{Zhao et al.}

\begin{abstract}
Vulnerability detection is garnering increasing attention in software engineering, since code vulnerabilities possibly pose significant security. Recently, reusing various code pre-trained models (e.g., CodeBERT, CodeT5, and CodeGen) has become common for code embedding without providing reasonable justifications in vulnerability detection. The premise for casually utilizing pre-trained models (PTMs) is that the code embeddings generated by different PTMs would generate a similar impact on the performance. \textbf{Is that TRUE?} To answer this important question, we systematically investigate the effects of code embedding generated by ten different code PTMs on the performance of vulnerability detection, and get the answer, i.e., \textbf{that is NOT true.} We observe that code embedding generated by various code PTMs can indeed influence the performance and selecting an embedding technique based on parameter scales and embedding dimension is not reliable. Our findings highlight the necessity of quantifying and evaluating the characteristics of code embedding generated by various code PTMs to understand the effects. To achieve this goal, we analyze the numerical representation and data distribution of code embedding generated by different PTMs to evaluate differences and characteristics. Based on these insights, we propose Coding-PTMs, a recommendation framework to assist engineers in selecting optimal code PTMs for their specific vulnerability detection tasks. Specifically, we define thirteen code embedding metrics across three dimensions (i.e., statistics, norm, and distribution) for constructing a specialized code PTM recommendation dataset. We then employ a Random Forest classifier to train a recommendation model and identify the optimal code PTMs from the candidate model zoo. We encourage engineers to use our Coding-PTMs to evaluate the characteristics of code embeddings generated by candidate code PTMs on the performance and recommend optimal code PTMs for code embedding in their vulnerability detection tasks. 

\end{abstract}



\received{31 May 2024}
\received[revised]{22 July 2024}
\received[accepted]{6 Aug 2024}

\maketitle

\section{Introduction}\label{sec:introduction}
Software vulnerabilities may lead to data leakage, information tampering, denial of service attacks (DoS), or other forms of security threats. Unfortunately, software vulnerabilities are inevitable due to software complexity and other reasons. Hence, software engineering researchers and practitioners are committed to proposing various detection methods, such as static analysis-based detection \cite{LippBP22}, machine learning-based detection \cite{ZhengGWLXLC20}, and deep learning-based ones \cite{0001DACW23}, to check whether there are vulnerability codes in software code snippets which may be exploited by attackers \cite{cheng-distributed-vd, LinWHZX20, HanifNRFA21, SenanayakeKAPP23}.

However, the code vulnerability detection method based on static analysis requires extra time to learn professional tools and it has a high false positive rate, while methods based on machine learning or deep learning require training a complete model from scratch, wasting too many resources. Currently with the popularity of pre-trained models (i.e., PTMs) trained on a wide range of datasets to grasp common features and have strong representation capabilities \cite{zhou-codebert4jit, tsabari2023predicting, akimova2021pytracebugs, tao2022c4}, especially the code pre-trained models \cite{kanade-cubert, feng-embedding, jiang-treebert, DevlinCLT19, guo-graphcodebert} have improved their code representation capabilities. Therefore, researchers in the code vulnerability detection field are also using code pre-trained models to perform code representation \cite{hoang-distributed-cc2vec, alon2019code2vec, hellendoorn-distributed-global} and then apply it to vulnerability detection to save training costs and improve performance.


Upon closer examination of recent research utilizing pre-trained contextual code embedding, we observed that these studies do not provide convincing reasons for choosing the corresponding specific code PTM to generate contextual embeddings for code snippets. 
For instance, Wang et.al \cite{WangXZX23} uses UniXcoder to capture the semantic relationship between programmatic nodes and their neighbors to obtain structural representations of the code snippets and then put them into the feed-forward network to predict a code snippet’s vulnerability and achieve excellent performance on traditional CWE datasets. While Zhang et.al \cite{ZhangLHXL23} adopted CodeBERT to learn the path representations of the control flow graph for the code snippet and used feature vectors as the representation of the code snippet and fed them into the classifier to detect vulnerabilities in ten datasets (e.g. CWE119, CWE20, CWE399) and also achieved superior performance. 
Nevertheless, none of them explained why they chose the specific code PTMs to generate code embedding representations nor compared the effects of using other code PTMs to generate code embeddings, although they all showed better results on partially the same datasets. 

Such previous research seems to use these code PTMs casually based on experience. However, the premise for casually utilizing code PTMs is that the code embeddings generated by different PTMs would generate a similar impact on vulnerability detection tasks, which has yet to be thoroughly investigated. Therefore, we first establish the following preliminary study by conducting experiments on ten different code embeddings generated by four different code PTMs covering all three architectures on four datasets for vulnerability detection tasks.

\textbf{RQ1: Whether code embeddings generated by different code PTMs affect the performance of vulnerability detection?} We observe that code embeddings generated by different code pre-trained models would result in different qualities and task performance. And larger parameter sizes code PTMs do not necessarily result in higher-dimensional code embeddings with better quality and task performance.

The experimental results of the preliminary study emphasize that code PTMs with larger parameter sizes cannot guarantee to generate high-dimensional code embeddings with better embedding quality and better task performance. Therefore, it is necessary to explore the characteristics and differences among code embeddings generated by different code pre-trained models on vulnerability detection tasks. So we further construct the next formative study by quantitative analysis of ten different code embeddings generated by four different code PTMs on four datasets for vulnerability detection tasks. 

\textbf{RQ2: What are the characteristics of the code embeddings generated by different PTMs?} We observe that different code PTMs yield code embeddings with varying numerical distributions, distinct numerical ranges, and distinct data distributions for the same dataset in the vulnerability detection task. Specifically, code embeddings generated by the CodeT5 family tend to exhibit the near-perfect normal distribution, while code embeddings from the PolyCoder family tend to manifest as skewed distributions.

Our preliminary study and formative study highlight that different pre-trained models generate code embeddings with distinct characteristics that significantly vary in embedding quality and result in diverse task performance. Moreover, selecting an embedding method solely based on the experience with parameter scale and embedding dimension is not reliable, as larger parameter scales and embedding dimensions do not guarantee results in higher-quality code embeddings and better task performance. Therefore, it is necessary to propose a recommendation framework to guide SE researchers and practitioners in selecting appropriate code PTM to generate high-quality code embeddings for better task performance.

Inspired by software metrics research and based on the results of the preliminary and formative study, we consider whether we can correlate the characteristics of code embeddings generated by different code PTMs with embedding quality and recommend embedding techniques based on these characteristics. Therefore, we define a set of thirteen code embedding metrics along three dimensions: statistics, norm, and distribution to quantify the differences and characteristics among different code embeddings and propose a recommended framework around these code embedding metrics to recommend optimal code pre-trained models for generating high-quality code embeddings with better task performance. Specifically, combining Devign, CWE119, CWE399 and Reveal four datasets of vulnerability detection and a total of four PTMs (i.e. CodeBERT \cite{feng-embedding}, CodeT5 \cite{wang-codet5}, PolyCoder \cite{xu2022systematic} and CodeGen \cite{nijkamp-codegen}) across three different architectures (i.e. encoder-only, decoder-only and encoder-decoder) for generating ten different dimensions embeddings, we constructed a new code embedding dataset that has thirteen dimensions of embedding metrics and a label indicating whether the embedding technique should be used. Based on this new dataset, we build a Random Forest (i.e., RF) classifier to train a recommendation model to determine whether the candidate code PTM generates high-quality code embeddings and would get better task performance.

We conduct experiments on the proposed framework on the constructed new dataset, and the results show that the defined code embedding metrics demonstrate a strong association with the embedding quality and the recommendation framework based on the metrics can achieve an Accuracy of 91\%, F1-Score of 83\%, AUC of 88\% and MCC of 77\%  to recommend appropriate code PTMs on the test sets. When applying the recommendation framework in practice, the PTMs recommended by the framework for the 78\% new unseen datasets were consistent with the PTMs that achieved the best performance. In addition, we provide practical guidance and clues for SE researchers and practitioners on how to use, iterate, and build our recommendation framework in the vulnerability detection task domain and other code-related task domains. 

In summary, the main contributions of this paper are as follows:
\begin{itemize}[topsep=1pt,itemsep=0.55pt]
    \item We systematically investigate the impact of code embeddings generated by different code PTMs on vulnerability detection tasks and analyze the differences and features in code embedding representations generated by code PTMs on code vulnerability detection tasks.
    \item We propose a set of thirteen metrics to quantify the differences and features among multiple code embeddings from three dimensions: statistics, norm, and distribution.
    \item We provide a recommendation framework based on the code embedding metrics to guide SE researchers and practitioners in choosing appropriate code PTM to generate high-quality code embeddings for better task performance.
\end{itemize}

\textbf{Paper Organization.} For the remainder of this paper, Section \ref{sec:related_works} discusses the related works. Section \ref{sec:study_design} presents the experimental datasets and code pre-trained models used in our study. Section \ref{sec:preliminary_study} and Section \ref{sec:formative_study} provide the motivation, approaches, and results in our preliminary and formative studies. Section \ref{sec:proposed_method} provides the details for constructing the recommendation framework. Section \ref{sec:assess} and \ref{sec:apply} evaluate the effectiveness of the framework in testing and practice. Section \ref{sec:discussion} provides practical guidance in using our recommendation framework. Section \ref{sec:threats_to_validity} discusses the threats to the validity of our results. Finally, Section \ref{sec:conclusions} concludes the paper.

\section{Related Work}\label{sec:related_works} 
This section introduces the related works of code embedding techniques and code vulnerability detection task.

\textbf{Code embedding techniques} aim to acquire compact code vector representations \cite{hoang-distributed-cc2vec, alon2019code2vec, hellendoorn-distributed-global}, referred to as code embeddings, to encapsulate the essence of source code, which have been instrumental in leveraging deep learning for tasks in software engineering (SE) \cite{siow-distributed-learning, kanade-cubert, watson2022systematic}. Many researchers have investigated the application of code embedding techniques in SE tasks. Kang et al. \cite{kang-assess} empirically evaluated the token embedding of Code2Vec \cite{alon2019code2vec} on three potential downstream tasks, namely code comment generation, code author identification, and code clone detection, and found that source code token embedding cannot be easily used for downstream tasks. Wang et al. \cite{wang2024empirical} evaluated the impact of five word embedding techniques, namely Word2Vec \cite{mikolov2013distributed}, GloVe \cite{pennington2014glove}, NextBug \cite{du2021deepsim}, ELMo \cite{peters2018dissecting}, and BERT \cite{DevlinCLT19}, on the bug assignment task and found that different word embedding models have a significant impact on the performance of deep learning-based bug assignment methods. Ding et al. \cite{ding-can} extended the experiments conducted by Kang et al. \cite{kang-assess} to six embedding techniques, namely Word2Vec \cite{mikolov2013distributed}, GloVe \cite{pennington2014glove}, FastText \cite{bojanowski2017enriching}, Code2Vec \cite{alon2019code2vec}, CuBERT \cite{kanade-cubert} and CodeBERT \cite{feng-embedding} and six downstream SE tasks, including code comment generation, code author identification, code clone detection, source code classification, log statement prediction, and software defect prediction. They found that using code embedding techniques does help achieve better performance in SE downstream tasks and that pre-trained contextual code embedding techniques such as CodeBERT and CuBERT outperform non-contextual code embedding techniques. Dou et al.\cite{dou2023towards} compared the performance of code embeddings of CodeBERT-base \cite{feng-embedding}, CodeBERT-MLM \cite{feng-embedding} and OpenAI's text-embedding-ada-002 in code clone detection and found that text-embedding-ada-002 provided the most robust performance in detecting cloned code pairs.

\textbf{Software vulnerability detection} is one of the methods to check and discover security vulnerabilities in software systems, and its purpose is to detect whether there are vulnerabilities in software code blocks that can be exploited by attackers. This task has been a research focus in the SE field for a long time. Since pre-trained contextual code embeddings facilitate downstream tasks, more and more SE researchers have begun to use pre-trained contextual code embeddings generated by different embedding techniques for code vulnerability detection tasks. For example, Sun et al.\cite{sun2023assbert} used the BERT model to obtain the feature representation of smart contract code, and then combined active learning technology and uncertain sampling strategy to learn contract vulnerability-related information from the feature representation, achieving good performance in contract vulnerability detection. Nguyen et.al \cite{NguyenNNLTP22} view each raw source code as a flat sequence of tokens to build a graph, wherein node features are initialized by the token embedding layer of CodeBERT and GraphCodeBERT and then leverage residual connection among GNN layers and obtain the highest accuracy on the real-world benchmark dataset from CodeXGLUE for vulnerability detection. Zhang et.al \cite{ZhangLHXL23} propose to decompose the syntax-based control flow graph of the code snippet into multiple execution paths and adopt CodeBERT to learn the path representations with intra- and inter-path attention. The feature vectors of the paths are combined as the representation of the code snippet and fed into the classifier to detect the vulnerability. 

Different from previous research on code embedding techniques and applying code PTMs to vulnerability detection tasks, our study focuses on systematically investigating the effects of different code embeddings on the vulnerability detection task, quantifying the characteristics and differences among these code embeddings and proposing a recommendation framework based on these characteristics to assist engineers in selecting optimal code PTMs for their specific vulnerability detection tasks.

\section{Experimental Data}\label{sec:study_design}
In this section, we provide a detailed description of the key components involved in the experiment of our study, including the PTMs utilized for generating code embeddings and the code vulnerability detection task datasets employed for experimental research.

\subsection{Code pre-trained models}
To ensure the comprehensiveness of the experiments in our study and the applicability of the conclusions, we examine four different code PTMs for generating ten different code embeddings. 
Specifically, we select CodeBERT, CodeGen, PolyCoder and CodeT5. As can be seen, a brief introduction to these four different models and the corresponding ten code embeddings is provided in Table \ref{tab:ptms}.

We choose the above code PTMs based on the following three reasons.
First, the selected code PTMs should be across all currently popular model architectures: encoder-only, encoder-decoder, and decoder-only. Therefore, we select CodeBERT for the encoder-only architecture, and for the decoder-only architecture, we employ CodeGen and PolyCoder. In the case of the encoder-decoder architecture, our choice is CodeT5. 
Second, each code PTM generates at least two embeddings of different sizes to allow only different embeddings to be compared under the same conditions.
Third, although the recent LLMs in the SE field are very popular and advanced, especially decoder-only architecture LLMs in code generation, such as CodeLlaMa. But the number of parameters of these models is extremely large, even the smallest version of CodeLlaMa has 7B parameters. So due to the limitations of hardware, we exclude these models. As an alternative, we select a small-scale version of the CodeGen series and PolyCoder family. These models also belong to the decoder architecture and have a good effect, so we use them to represent other LLMs of the same decoder-only architecture.

\begin{table}[!ht]
\vspace{-0.25cm}
  \caption{Ten different code embeddings of four code PTMs.}
  \label{tab:ptms}
  \begin{tabular}{p{2.2cm}p{2.3cm}p{1.5cm}p{1.5cm}}
    \toprule
    PTM Name  &Architecture& Size  & Dimension \\
    \midrule
   CodeBERT-small &Encoder&84M&768\\
   CodeBERT-base &Encoder&125M&768\\   
   CodeGen-small &Decoder&350M&1024\\
   CodeGen-base &Decoder&2B&2560\\    
   CodeT5-small &Encoder-Decoder&60M&512\\
   CodeT5-base &Encoder-Decoder&220M&768\\
   CodeT5-large &Encoder-Decoder&770M&1024\\
   PolyCoder-small &Decoder&160M&768\\
   PolyCoder-base &Decoder&0.4B&1024\\
   PolyCoder-large &Decoder&2.7B&2560\\

  \bottomrule
\end{tabular}
\vspace{-0.4cm}
\end{table}

\subsection{Vulnerability detection datasets}

Benefiting from the fact that vulnerability detection tasks have long been a research focus in the SE field, relevant datasets can be easily obtained and used out of the box. Therefore, we collected the datasets needed in the open-source community and we easily collected the datasets for the vulnerability detection tasks thanks to the spirit of open sharing. As can be seen, Table \ref{tab:dataset} presents brief information on the studied datasets. We use four datasets including Devign, Reveal, CWE119, and CWE399, where Reveal and Devign are vulnerability datasets composed entirely of real-world exploit codes collected by Chakraborty et al. \cite{Reveal} and Zhou et al. \cite{zhou-devign}, respectively. CWE119 and CWE399 are vulnerability datasets collected by Li et al. \cite{li-cvd-vuldeepecker}, which contain some synthetic examples. 

\begin{table}[!ht]
\vspace{-0.3cm}
  \caption{Statistics for vulnerability detection tasks dataset.}
  \label{tab:freq}
  \begin{tabular}{ccccccc}
    \toprule
Project    & Examples & \%Vulnerable ratio & Language  \\
    \midrule
CWE119    & 39753            & 26.26                 & C/C++    \\
CWE399    & 21885            & 33.29                 & C/C++    \\
Reveal    & 18169            & 9.16                  & C/C++    \\
Devign    & 22361            & 45.02                 & C/C++   \\
  \bottomrule
\end{tabular}
\label{tab:dataset}
\vspace{-0.35cm}
\end{table}

\begin{figure*}[!tbph]
	\centering
	\subfigure{
		\label{fig:rq1-devign}
\includegraphics[width=0.48\linewidth,height=0.20\textheight]{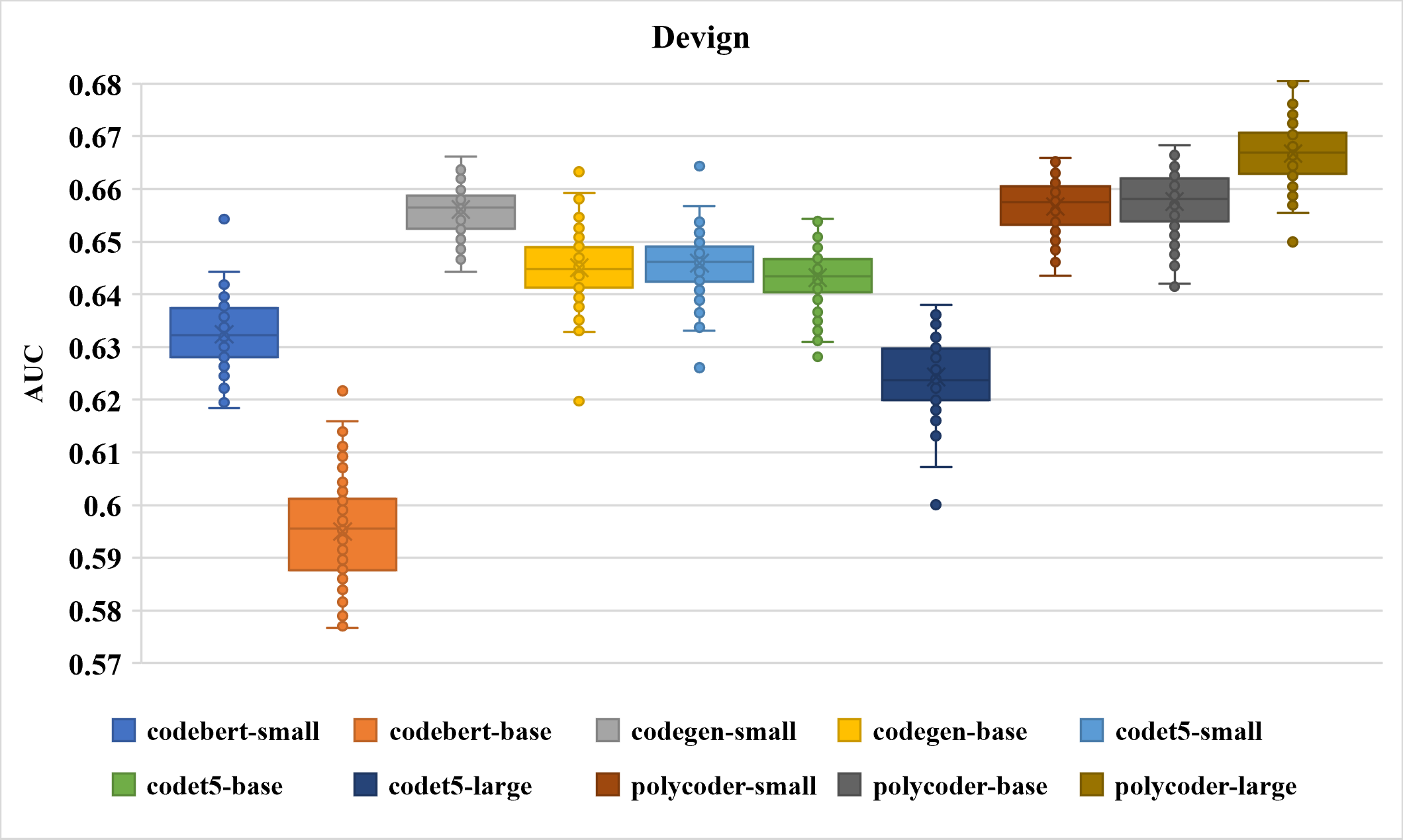}}
    \subfigure{
		\label{fig:rq1-reveal}
\includegraphics[width=0.48\linewidth,height=0.20\textheight]{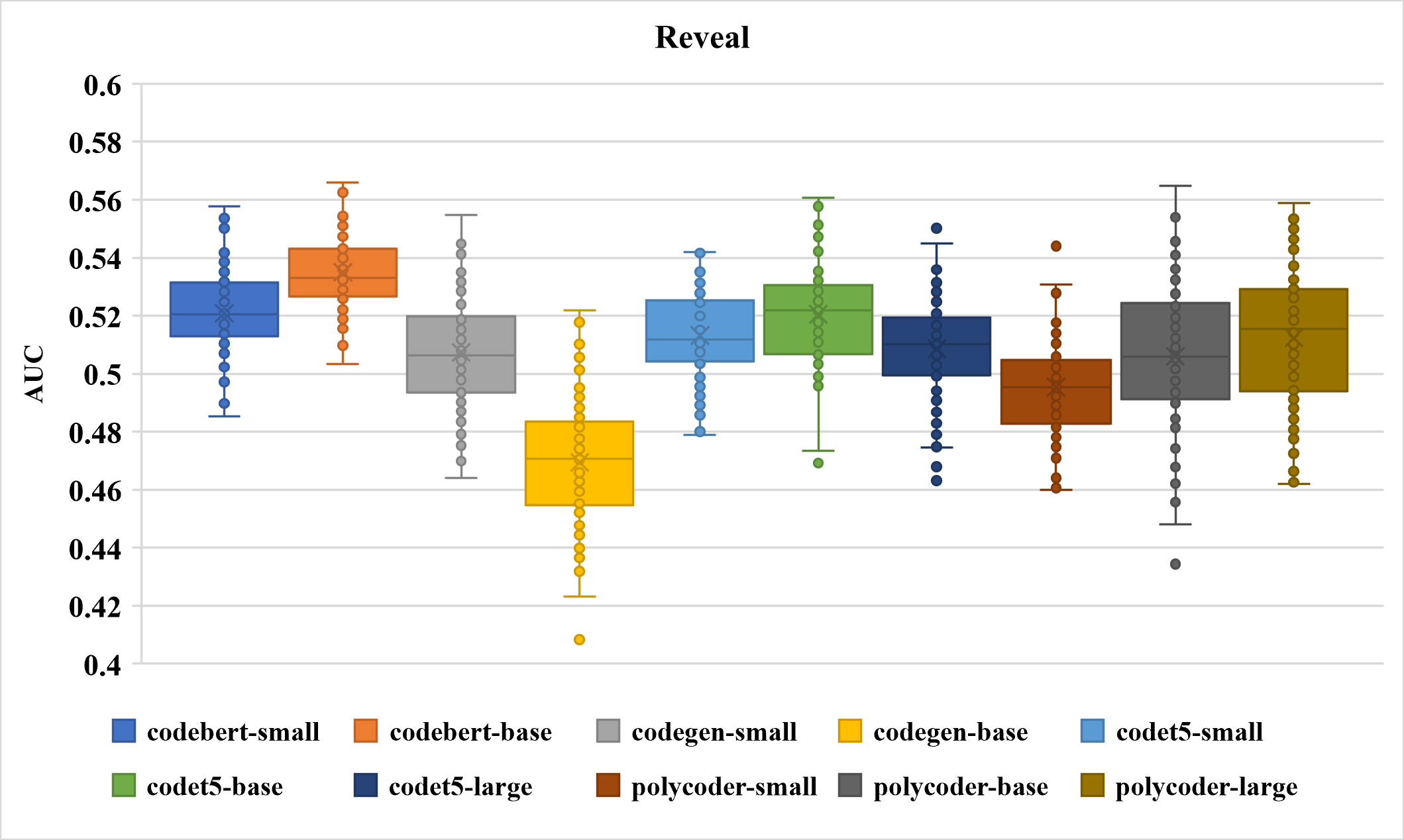}}
	\subfigure{
		\label{fig:rq1-cwe119}
\includegraphics[width=0.48\linewidth,height=0.20\textheight]{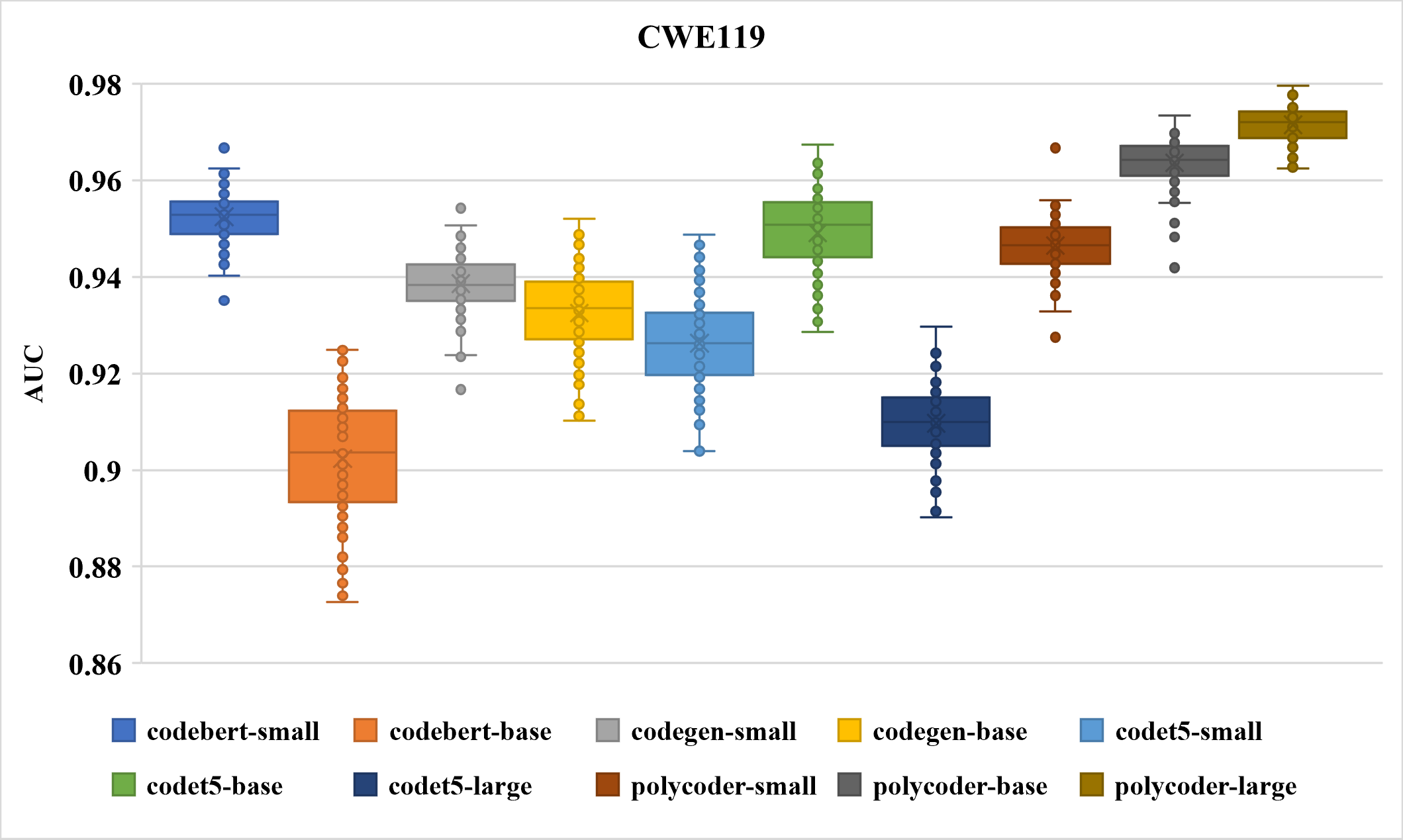}}
    \subfigure{
		\label{fig:rq1-cwe399}
\includegraphics[width=0.48\linewidth,height=0.20\textheight]{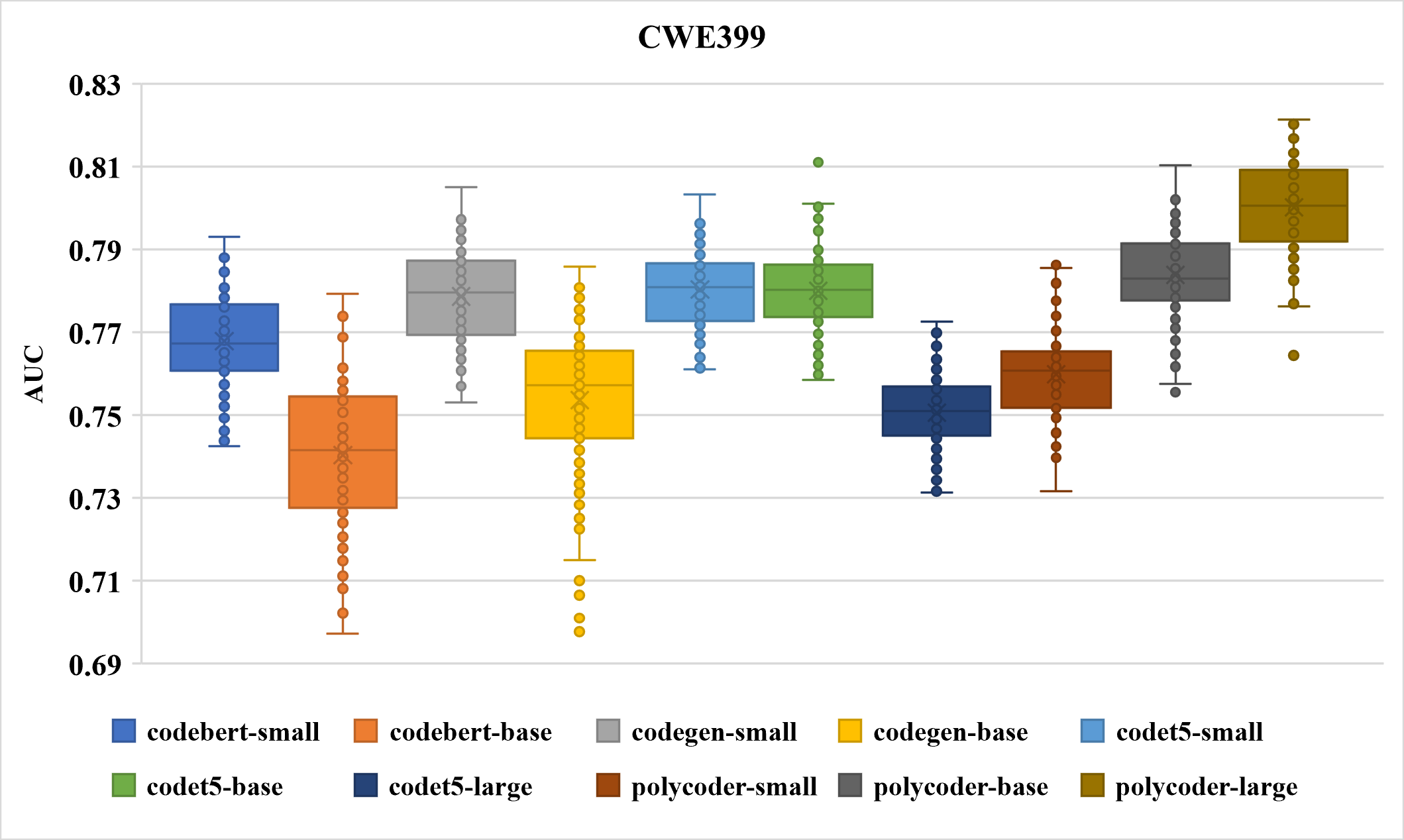}}
   \caption{Performance distribution of the AUC values obtained on the corresponding test data of the classifier built based on ten different code embeddings generated by four different code PTMs on the four datasets of the vulnerability detection task.}
   \label{fig:rq1}
   
\vspace{-0.45cm}
\end{figure*}

\section{Preliminary Study}\label{sec:preliminary_study}
The current mainstream for researchers and practitioners is to apply code PTMs spanning various programming languages, diverse architectures, and various application scenarios to code vulnerability detection tasks, aiming to generate preliminary code embeddings for code snippets potentially harboring vulnerabilities and then integrate them into the pipeline of task-specific deep learning models to save training costs and improve performance. However, researchers casually use various code PTMs to generate contextual embeddings without providing convincing reasons for choosing the corresponding specific code PTMs and also have no idea about the impact of code embeddings generated by different code PTMs on vulnerability detection tasks. This may result in using an inappropriate code PTM to generate code embeddings with lower quality and thus achieve poorer performance. Hence, 
in this section, we construct a preliminary study to understand the impact of code embeddings generated by different code PTMs on vulnerability detection tasks. In more detail, we will proceed by answering the following question:


\textbf{RQ1: Whether code embeddings generated by different code PTMs affect the performance of vulnerability detection?}


\textbf{Approach:} 
In order to tackle this issue and ensure the accuracy of our conclusions, we conducted experiments on ten distinct code embeddings generated by the four different code PTMs covering all three architectures chosen in this study. The experiments are conducted on four frequently used datasets covering vulnerability detection tasks \cite{ZhouLSD019, NguyenNNLTP22, ChakrabortyKDR22, TangTBZF23}.

To assess the quality of code embeddings generated by different PTMs, akin to prior methodologies \cite{sharma-attention-bert, ding-can, wang2024empirical, tenney2019bert}, we construct and train a simple fully connected layer based on the acquired code embeddings from input code snippets for classification. The classifier head over the code embeddings remains fine-tuned on the task training data. We gauge the performance of classifiers on test data to discern the quality of code embeddings produced by various code PTMs. Specifically, we compare the AUC values of the classifiers for discrimination, as the AUC value is a threshold-independent metric less susceptible to dataset class imbalance. Given that this method doesn't entail fine-tuning the parameters of the pre-trained model, it solely generates initial code embeddings for code snippets. Prior studies have indicated that averaging the vector representations of all code tokens to obtain the code embedding of the entire input can yield higher-quality embeddings compared to utilizing special tokens. Therefore, we adopt the approach of average pooling across all input code tokens excluding special tokens to get code embeddings.

In order to ensure the fairness of the experiment, the input information of each model is the same, that is, the length of the input code is controlled at 100 tokens. To mitigate performance biases resulting from experimental randomness, we conducted the experiment 100 times and then compared the distribution of performance values. 

\textbf{Results:} Figure \ref{fig:rq1} shows the distribution of AUC performance values obtained on the corresponding test sets for classifiers built based on the different code embeddings generated by code PTMs on the four datasets of the vulnerability detection task.

\textbf{Observation 1) Under the same parameter scale, code embeddings of approximately equal dimensions, generated by different code PTMs, display notable discrepancies in their semantic content and task performances.} We categorize the ten different code embeddings listed in Table \ref{tab:ptms} into four groups based on the parameter scales of the code PTMs. These groups consist of code embeddings generated by CodeBERT-small and CodeT5-small with parameter sizes less than 100M, CodeGen-base and PolyCoder-large with parameter sizes exceeding 2B, CodeBERT-base, CodeT5-base and PolyCoder-small with parameter sizes between 100M and 300M, as well as CodeGen-small, CodeT5-large and PolyCoder-base with parameter sizes ranging from 300M to 1B. With the exception of the code embedding dimensions generated by CodeBERT-small and CodeT5-small, which are 768 dimensions and 512 dimensions, respectively, the code embedding dimensions generated by each group of PTMs with the same scale are largely consistent. For instance, the code embedding dimensions produced by PTMs with parameters exceeding 2B are all 2560 dimensions, while those generated by the remaining two groups are 768 dimensions and 1024 dimensions, respectively. Further analysis of Figure \ref{fig:rq1} reveals significant variations in the performance of classifiers constructed using code embeddings from different code PTMs within each group sharing the same parameter scale. For instance, across all four datasets, classifiers built on code embeddings from PolyCoder-large consistently outperform those built on code embeddings from CodeGen-base. Additionally, while the classifier performance based on code embeddings from CodeBERT-small surpasses that from CodeT5-small on the Reveal and CWE119 datasets, the reverse is observed for the Devign and CWE399 datasets. \textbf{This observation indicates that initial code embeddings generated by different code PTMs for vulnerability detection tasks under the same parameter scale harbor varying degrees of semantic richness and notable discrepancies in embedding quality and task performances. This observation holds even when not restricted to the same parameter scale.} Thus, the question of how to select the appropriate code PTM from these candidate PTMs to produce high-quality initial code embeddings for vulnerability detection tasks warrants further exploration.

\textbf{Observation 2) Within the same model family, the initial code embedding dimensions generally increase as the parameter scale of different code PTMs for vulnerability detection tasks increases. However, larger parameter sizes in code PTMs do not necessarily result in higher-dimensional code embeddings containing richer semantic information and better task performance. Sometimes smaller PTMs produce higher-quality code embeddings and better performance.} As depicted in Table \ref{tab:ptms} and Figure \ref{fig:rq1}, the initial code embedding dimensions generated by the CodeGen family, CodeT5 family, and PolyCoder family for vulnerability detection tasks all increase with the PTM parameter size escalation excluding the CodeBERT family. However, the performance of classifiers constructed using higher-dimensional code embeddings from code PTMs with larger parameter sizes is not guaranteed to be improved. For instance, even though the parameter size of CodeGen-small is 350M, yielding an initial code embedding dimension of 1024 dimensions for the vulnerability detection task, its classifier's performance notably surpasses that of the classifier built on the 2560-dimensional code embedding from the 2B parameter CodeGen-base across all four datasets. Similarly, the performance of the classifier constructed using the code embedding from CodeT5-large is notably inferior to those built using CodeT5-small and CodeT5-base across all four datasets. Remarkably, only the performance of classifiers constructed using the code embeddings from the three polycoder PTMs on the Reveal, CWE119, and CWE399 datasets adheres to the expected rule that larger PTM scales yield higher-dimensional embeddings with better classifier performance. However, on the Devign dataset, the performance of the classifier constructed using the code embedding from polycoder-base is comparable to that of the classifier built using the code embedding from PolyCoder-small. \textbf{This phenomenon underscores the absence of a guarantee that code PTMs with larger parameter sizes will generate higher-dimensional code embeddings with richer semantic information and superior embedding quality with better task performance. Conversely, smaller models may indeed produce higher-quality code embeddings and better performance. Moreover, this trend persists across multiple model families.} For example, on the Reveal dataset, the performance of classifiers constructed using code embeddings from CodeBERT-small and CodeBERT-base with smaller parameter sizes surpassed those of the remaining code PTM family members with larger parameter sizes.

\textbf{Summary.} Through the experimental results of our preliminary study, we observe that different pre-trained models generate code embeddings that significantly vary in quality and task performance. And cannot guarantee that code PTMs with larger parameter sizes will generate high-dimensional code embeddings with better embedding quality and better task performance. Therefore, it is necessary to explore the characteristics and differences between code embeddings generated by different pre-trained models on vulnerability detection tasks.

\section{Formative Study}\label{sec:formative_study}
As evidenced by the results of RQ1, different code PTMs yield code embeddings of significantly divergent quality and task performance for code vulnerability detection. Yet, the underlying principles driving these phenomena remain largely unexplored. This situation inspires us to consider: given the challenges in unifying the styles of these code PTMs (Such as differentiated pre-training tasks), why not shift our focus to understanding and analyzing the characteristics and differences of these code embeddings? Therefore,
in this section, we construct a formative study to explore, evaluate, and quantify the characteristics and differences among different code embeddings generated by various code PTMs. Specifically, we will continue to answer the following question to move forward:

\textbf{RQ2: What are the characteristics of the code embeddings generated by different PTMs? }

\begin{figure*}[!tbph]
	\centering
	\subfigure{
		\label{fig:rq2-1-train}	\includegraphics[width=0.48\linewidth,height=0.18\textheight]{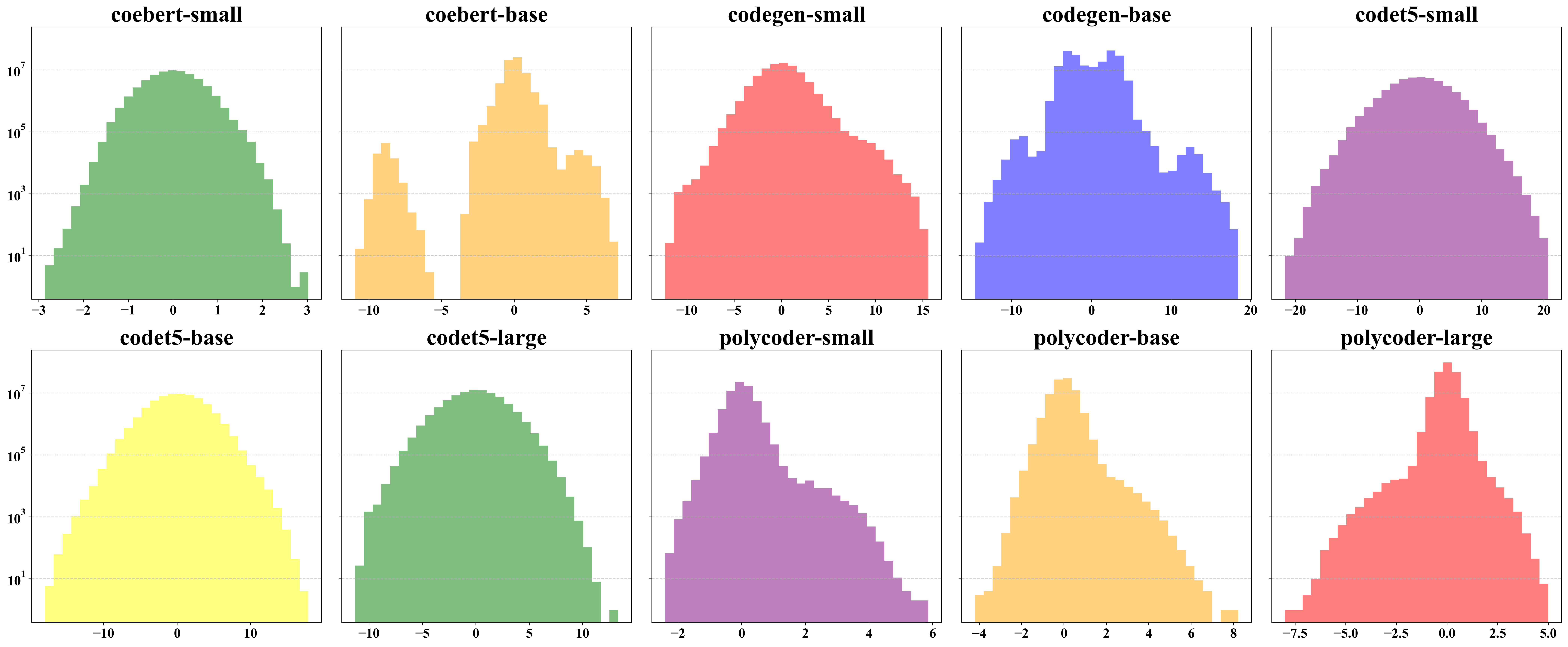}}
    \subfigure{
		\label{fig:rq2-1-test}	\includegraphics[width=0.48\linewidth,height=0.18\textheight]{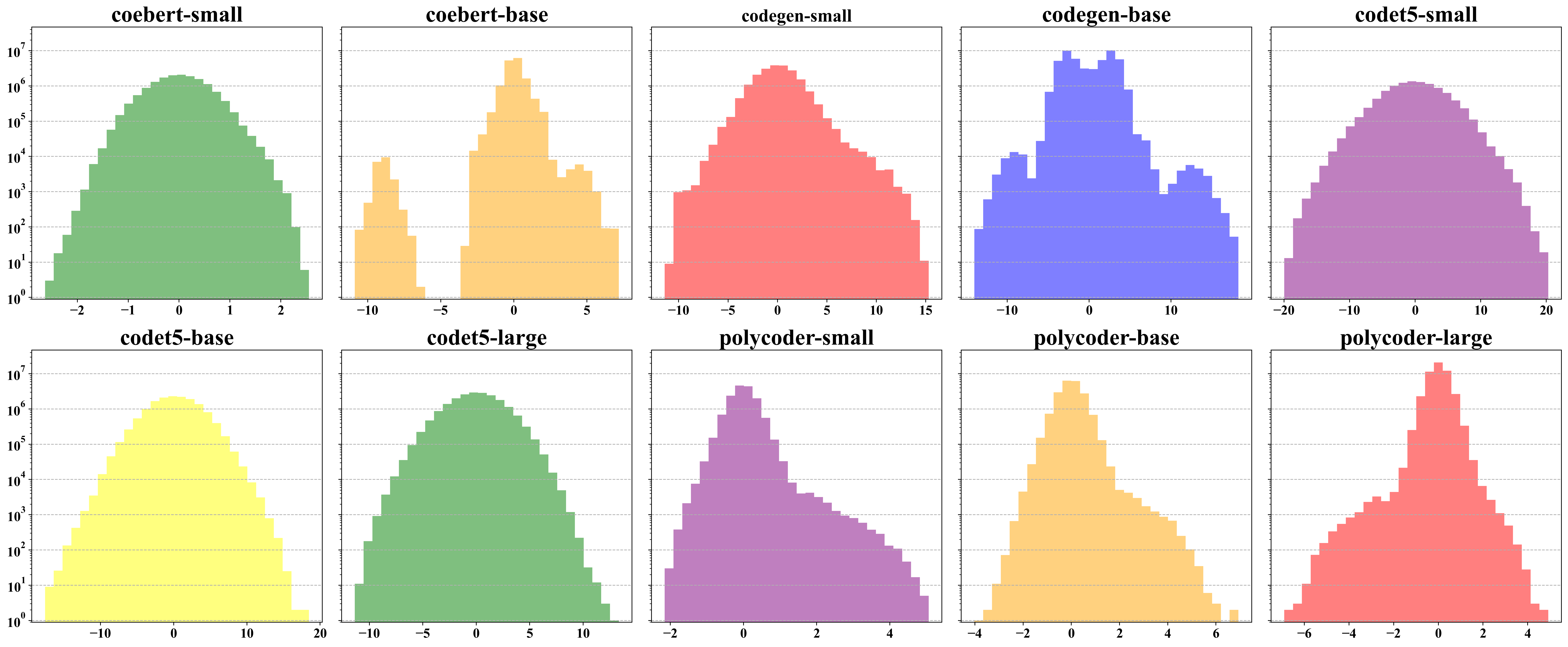}}
   \caption{Numerical distributions of the code embeddings generated by different PTMs on vulnerability detection task, where the abscissa represents the value of the code embeddings, and the ordinate represents the frequency of the corresponding value, which is represented in logarithmic form.}
   \label{fig:rq2-1}
\vspace{-0.3cm}
\end{figure*}

\begin{figure*}[!tbph]
	\centering
	\subfigure{
		\label{fig:rq2-2-train}
\includegraphics[width=0.48\linewidth,height=0.18\textheight]{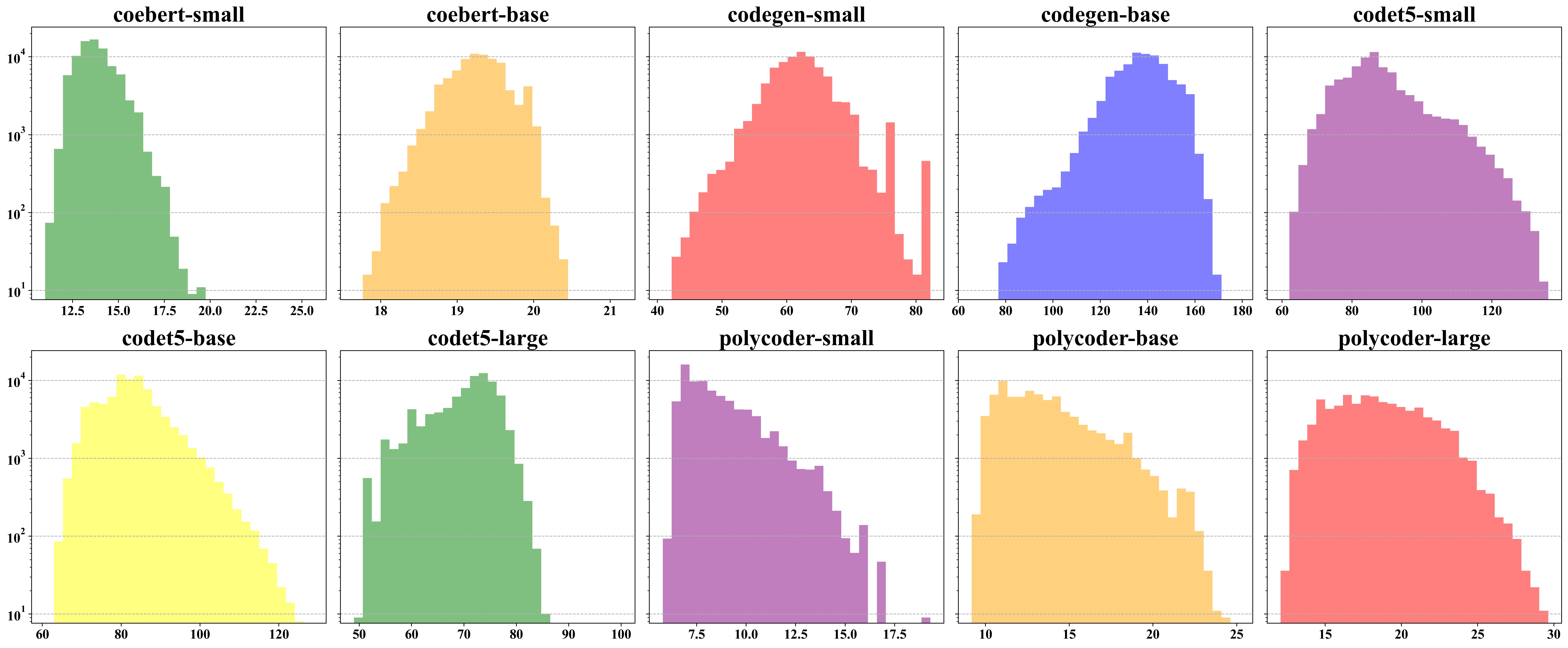}}
    \subfigure{
		\label{fig:rq2-2-test}
\includegraphics[width=0.48\linewidth,height=0.18\textheight]{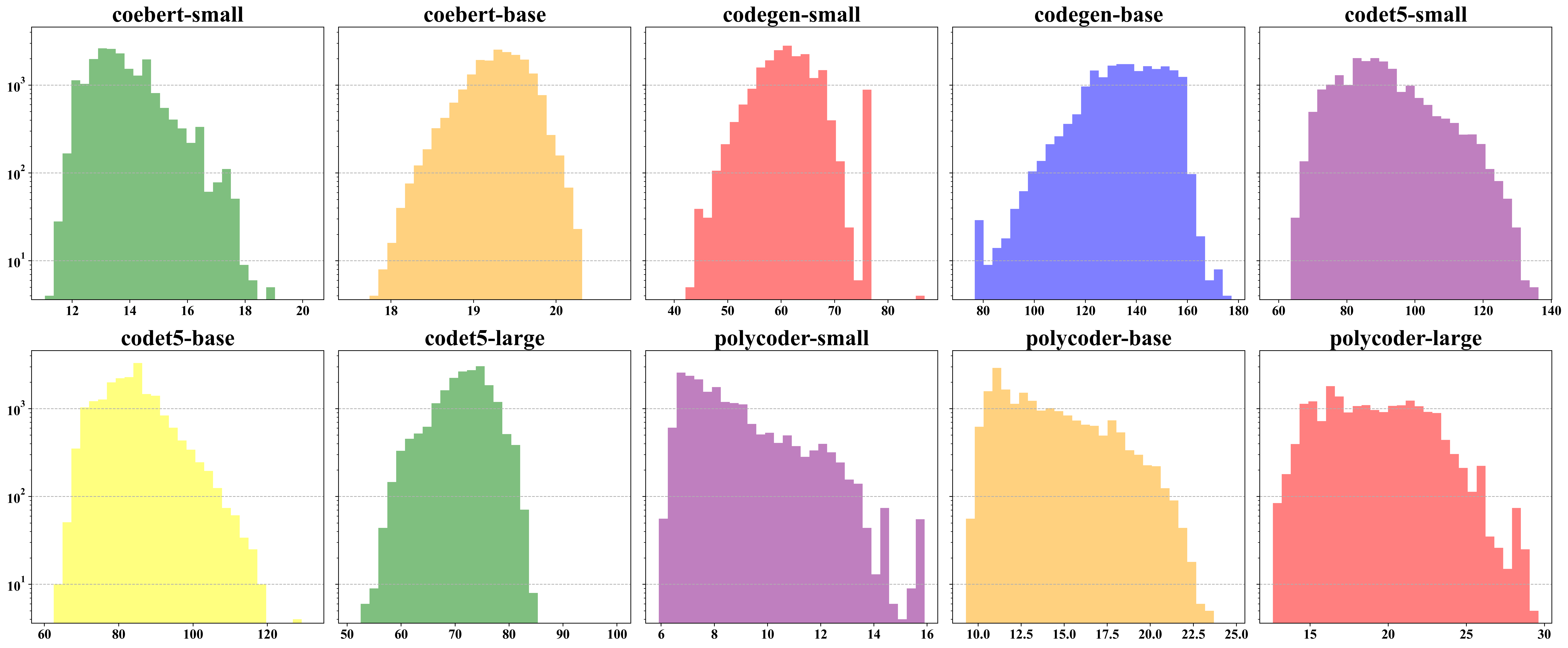}}
   \caption{Data distributions of the code embeddings generated by different PTMs on vulnerability detection task, where the abscissa represents vector L2 norm's value, and the ordinate represents the frequency of the corresponding value, which is represented in logarithmic form.}
   \label{fig:rq2-2}
\vspace{-0.45cm}
\end{figure*}


\textbf{Approach:} 
To address this RQ, we begin by computing the code embeddings generated by the code PTMs under study for the four datasets of the vulnerability detection task. Subsequently, for clarity in the presentation, we merge the training sets and the test sets of the four datasets separately, resulting in combined code embeddings for training data and test data. Finally, we visually represent these code embeddings to observe their characteristics. Specifically, we visualize the code embeddings generated by various code pre-trained models for both the vulnerability detection training data and test data to discern differences in numerical range, numerical distribution, and data distribution among them. To analyze the numerical range and numerical distribution of code embeddings, we flatten the code embedding matrices produced by different PTMs and conduct statistical analyses on the distribution of values. For data distribution, we compute the L2 norm of the vector representation of each instance in the code embedding representation of the dataset. Subsequently, we employ the L2 norm distribution of the vector representation of all instances of code embedding to depict the data distribution of the entire code embedding. Note that the method used to generate code embeddings remains consistent with that of RQ1, involving the average pooling of vector representations of all code tokens to obtain the code embedding for the entire input while controlling the input information of each model to consist of the same 100 tokens.

\textbf{Results:} Figure \ref{fig:rq2-1} shows numerical distributions of the embedding representations generated by different PTMs on the vulnerability detection task. Figure \ref{fig:rq2-2} shows the data distributions of the embedding representations generated by different PTMs on the vulnerability detection task. The left of the two figures is the code embedding generated using the training dataset, and the right is the code embedding generated using the test dataset.

\textbf{Observation 3) In the vulnerability detection task, PTMs from different families yield code embedding representations with distinct numerical ranges. Moreover, even within the same family, PTMs with varying parameter sizes generate code embedding representations with differing numerical ranges.} As illustrated in Figure \ref{fig:rq2-1}, on the training dataset of the vulnerability detection task, code embeddings generated by CodeT5-small and CodeT5-base exhibit the widest value range. Both embeddings yield similar value ranges, with embedding values typically falling between -20 and 20. Following closely is CodeGen-base, with generated code embedding values ranging between -15 and 20. Code embeddings from CodeT5-large and CodeGen-small generally range between -15 and 15. In contrast, the numerical values represented by code embeddings from the CodeBERT and PolyCoder families are relatively small, with numerical ranges not exceeding 20. Notably, CodeBERT-small exhibits the smallest numerical range among the ten code PTMs, with code embedding values primarily ranging between -3 and 3. On the other hand, CodeBERT-base, featuring a larger parameter scale within the same family, yields code embedding values ranging roughly between -10 and 10. Similarly, code embedding values from the PolyCoder family's PTMs with three different parameter sizes fall between -2 and 6, -4 and 8, and -7.5 and 5, respectively. Similar phenomena are observed on the test dataset of the vulnerability detection task. 

\textbf{Observation 4) In the vulnerability detection task, distinct PTMs yield code embeddings with varying numerical distributions. Specifically, code embeddings generated by the CodeT5 family tend to exhibit the near-perfect normal distribution. Meanwhile, those from the CodeBERT and CodeGen families also tend towards the normal distribution but display certain imperfections. On the other hand, code embeddings from the PolyCoder family tend to manifest as skewed distributions.} As depicted in Figure \ref{fig:rq2-1}, code embeddings generated by CodeT5-small, CodeT5-base, and CodeT5-large from the CodeT5 family tend towards a normal distribution, with numerical values symmetrically distributed around 0. Similarly, the numerical distribution of code embeddings from CodeBERT-small also leans towards a normal distribution, albeit with slight deviations, particularly in the range between 2 and 3 on the right side. This pattern is observed in CodeGen-small and CodeGen-base as well. Although the predominant representation of these embedded values comprises floating-point numbers distributed around 0, forming a trend towards normal distribution, there remains a notable number of values dispersed around other values, such as those around -10 and 10. Conversely, code embeddings generated by the PolyCoder family's PolyCoder-small, PolyCoder-base, and PolyCoder-large exhibit skewed numerical distributions. Specifically, those from PolyCoder-small and PolyCoder-base display positive skewness, with a concentration of values between 2 and 4. However, code embeddings from PolyCoder-large demonstrate negative skewness, with certain values concentrated between -5.0 and -2.5. Unlike the continuous distribution observed in code embeddings generated by the nine aforementioned PTMs, the distribution of embedding representations produced by CodeBERT-base is discrete. These discrete segments of values each tend towards a normal distribution. Similar patterns are also observed in the test dataset of the vulnerability detection task. 

\textbf{Observation 5) In the vulnerability detection task, different PTMs yield code embeddings with distinct data distributions for the same dataset. While the data distributions of code embeddings between the training dataset and the test dataset generally appear similar, subtle differences can also be observed.} As illustrated in Figure \ref{fig:rq2-2}, significant disparities exist in the data distribution of code embeddings generated by each code PTM within the same family and across different families. Notable examples include CodeBERT-small and CodeBERT-base, where the former exhibits a tendency towards a positively skewed distribution, while the latter tends towards a negatively skewed distribution. Similar patterns are observed with CodeGen-small and CodeGen-base, where the former's data distribution tends towards a flawed yet normal distribution, while the latter leans towards a negatively skewed distribution. Although the data distributions of code embeddings generated by the three PTMs of the PolyCoder family all skew to the left, substantial differences in values and peak values of the vector L2 norm within the code embeddings are evident. Furthermore, we observe a general similarity in the shape of the data distribution of code embeddings between the training dataset and the test dataset. For instance, the data distribution of code embeddings generated by CodeGen-small for the training dataset and the test dataset both tends towards a normal distribution around the vector L2 norm value of 60. However, differences in values on both sides are noticeable. This suggests that the diverse code embeddings generated by code PTMs for the training dataset and test dataset capture the distribution disparities between the two. Existing research indicates that deep learning models can adeptly learn relevant knowledge, such as code structure information, when trained on datasets with specific data distributions, such as a normal distribution. Therefore, the significant divergence in the data distribution of code embeddings within the same dataset implies variations in the code semantics and structural information encapsulated within these different code embeddings. Additionally, code embeddings can also capture patterns and disparities between different data distributions across training and test datasets.

\textbf{Summary.} Through the quantitative results of our preliminary study and formative study, we observe that different pre-trained models generate code embeddings with distinct characteristics that significantly vary in semantic quality. Moreover, selecting an embedding method solely based on the experience with parameter scale and embedding dimension is not reliable, as larger parameter scales and embedding dimensions do not guarantee results in higher-quality code embeddings. Therefore, it is necessary to propose a recommendation framework to guide SE researchers and practitioners in selecting appropriate code PTM to generate high-quality code embeddings. 

\section{Proposed Framework to Recommend Optimal Code Pre-trained Models}\label{sec:proposed_method}

\begin{figure*}

  \centering
  \includegraphics[width=1.0\linewidth,height=0.22\textheight]{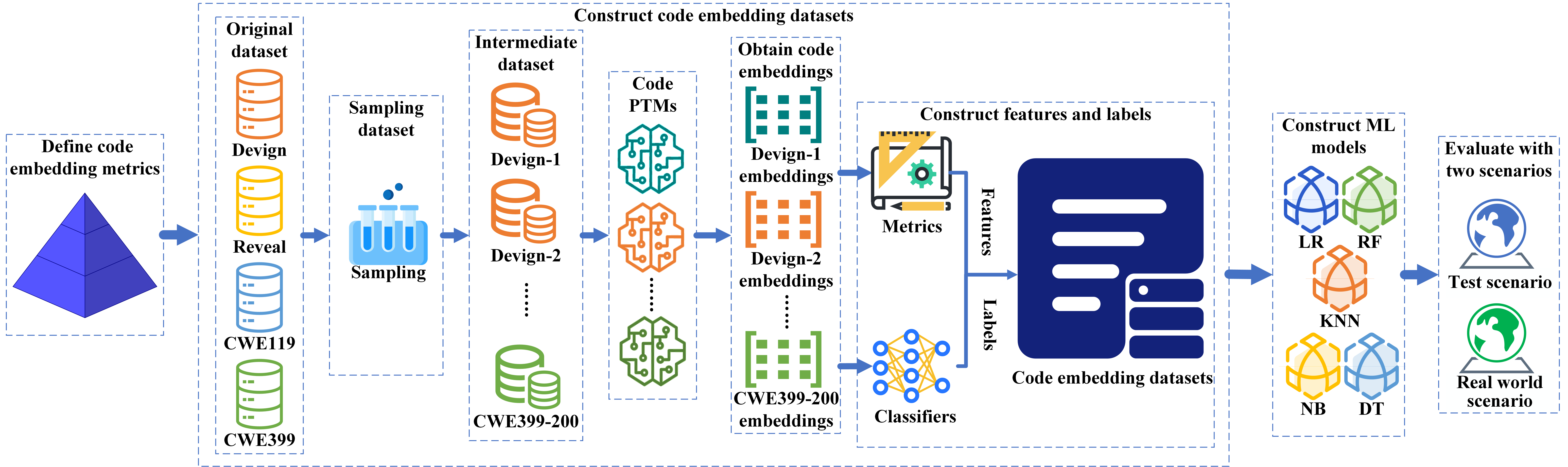}
  \caption{Recommendation framework.}
\label{fig:rq3}
\vspace{-0.25cm}
\end{figure*}

The preliminary and formative studies prompt us to consider whether we can correlate the characteristics of code embeddings generated by different code PTMs with embedding quality and recommend embedding techniques based on these characteristics. Therefore, we propose a recommended framework around code embedding metrics to recommend optimal code pre-trained models for generating code embeddings outlined in Figure \ref{fig:rq3} and we will elaborate on this framework in detail, following a series of steps.

\begin{table*}[!ht]
  \caption{Description of the defined metrics for code embeddings.}
  \label{tab:metrics}
  \small
\begin{tabular}{|c|l|c|}
\hline
\textbf{Name}                     & \textbf{Description}                                                                                                        & \textbf{Dimension}                  \\ \hline
Range                    & The difference value between the maximum numerical value and the minimum numerical value in the code embedding     & \multirow{7}{*}{Statistic} \\ \cline{1-2}
Mean                     & The average value of all numerical values in the code embedding                                                    &                            \\ \cline{1-2}
Median                   & The middle value of all numerical values in the code embedding                                                     &                            \\ \cline{1-2}
Var                 & A measure of how far numerical values are spread out from the average value in the code embedding                  &                            \\ \cline{1-2}
Std       & The square root of the variance                                                                                    &                            \\ \cline{1-2}
Skew                 & The numerical characteristics of the degree of asymmetry of data distribution in the code embedding                &                            \\ \cline{1-2}
Kurt                 & The numerical characteristics of morphological steepness and gentleness of data distribution in the code embedding &                            \\ \hline
L0 Norm                  & The number of non-zero elements in the code embedding                                                              & \multirow{4}{*}{Norm}      \\ \cline{1-2}
L1 Norm                  & The sum of the absolute values of each element in the code embedding                                               &                            \\ \cline{1-2}
L2 Norm                  & The square root of the sum of the squares of the elements in the code embedding                                    &                            \\ \cline{1-2}
Nuclear Norm             & The maximum value of the sum of absolute values of row vectors in the code embedding                               &                            \\ \hline
MMD & The kernel distance between the data distributions corresponding to code embeddings                                & \multirow{2}{*}{Distribution}  \\ \cline{1-2}
Dot Product              & The product of the length and the cosine of the angle between vectors-based corresponding to code embedding        &                            \\ \hline
\end{tabular}
\vspace{-0.4cm}
\end{table*}

\subsection{Define code embedding metrics}
Since software metrics serve as measurements of software properties and specifications, aiming to provide objective, replicable, and quantifiable results to aid in software development and maintenance. For instance, research in software defect prediction utilizes metrics like the classic CK metric \cite{ChidamberK94CK, TangKC99} to gauge software complexity. These metrics are then employed to construct software defect prediction machine learning models for predicting whether software modules contain potential defects, thereby aiding in the rational allocation of limited testing resources \cite{ZhangHMZ17, HeLLCM15, LiuLGZX18}. Inspired by relevant research in the aforementioned fields and formative study prompts us to recognize the potential to leverage the distinct characteristics of code embeddings identified in RQ2 to design a set of code embedding metrics for measuring the quality of code embeddings. 

Therefore, we define a set of thirteen metrics to quantify the differences and characteristics of code embeddings as observed in RQ2. As outlined in Table \ref{tab:metrics}, we measure the characteristics of code embeddings from three perspectives, which correspond to three observations in RQ2. 

\textit{Statistical metrics: }This encompasses metrics such as Mean, Median, Variance (Var), Standard Deviation (Std), Skewness (Skew), Kurtosis (Kurt), and Range of code embeddings. These metrics correspond to the first and second observations in RQ2. For instance, Range quantifies the span of values, while statistical features like Skewness and Kurtosis enable the measurement of different numerical distributions.

\textit{Normative metrics:} This corresponds to the third observation of RQ2 and is used to assess the data distribution of code embeddings. It includes metrics such as the L2 norm, Nuclear norm, L0 norm, and L1 norm of code embeddings. These norms, commonly used in NLP, gauge sentence difficulty in word embeddings, word information gain, and word importance \cite{ LiuLWC20, OyamaYS23, YokoiTASI20}. 

\textit{Distribution metrics:} This also corresponds to the third observation of RQ2, focusing on similarities and disparities between the data distributions of code embeddings in the training set and the test set. To measure this, we adopt Maximum Mean Discrepancy (MMD) distance and dot product distance. MMD distance is a non-parametric measure based on the kernel method, capable of effectively gauging similarities or differences between samples without relying on distribution assumptions. Dot product distance is a simple and efficient measurement method with an intuitive geometric interpretation, which considers both numerical values and angles simultaneously. These two distances are complementary in this study, with dot product distance capturing data distributions with strong linear correlation in code embeddings, while MMD is adept at capturing complex nonlinear data distributions.

\subsection{Construct code embedding datasets}
Subsequently, we constructed a new dataset called the code embedding dataset derived from the existing vulnerability dataset. The distinguishing feature of this new dataset lies in the incorporation of the thirteen code embedding metrics defined in Table \ref{tab:metrics}. The labels assigned to instances in this new dataset are binary, indicating whether the embedding is deemed to be of high quality or not. The construction details of the code embedding dataset are as follows.

\textit{Sampling original datasets:} For the above four vulnerability datasets, we randomly sampled each dataset 200 times while ensuring the original data distribution, and obtained 800 new datasets. 

\textit{Obtain code embeddings:} For these 800 new datasets, we use our four different code PTMs to generate ten different candidate code embeddings and finally we obtain 8000 code embeddings. Note that the method used to generate code embeddings remains consistent with that of RQ1 and RQ2, involving the average pooling of vector representations of all code tokens to obtain the code embedding for the entire input while controlling the input information of each model to consist of the same 100 tokens.

\textit{Construct features and labels:} Then based on the code embedding metrics defined and presented in Table \ref{tab:metrics}, we compute and generate a new code embedding dataset, comprising 13-dimensional metrics features. Subsequently, we proceed to create labels for this dataset. Similar to the experimental process of RQ1, in order to compare the quality of code embeddings generated by different PTMs, we build and train a simple fully connected layer for classification based on the obtained code embeddings. The head of the classifier is still fine-tuned on the training data. We compare the performance of classifiers on test data to determine the quality of code embeddings generated by different code PTMs. In this way, the ranking of the quality of 10 code embeddings on the same dataset is reflected by the AUC value of the classifier. We mark the code embeddings corresponding to the three code PTMs with the top three AUC values as positive classes, and the code embeddings corresponding to the last seven code PTMs as negative classes. Finally, a code embedding dataset containing 8,000 instances and 13 feature dimensions was formed. 

\subsection{Construct machine learning models for recommendation}
By constructing the code embedding dataset, we transform the code PTMs recommendation problem into a binary classification decision problem. That is, recommending an appropriate code PTM turns into a binary classification problem of whether to use the code embeddings generated by the specific code PTM. To achieve this goal, we build the traditional machine learning models by using code embedding metrics and labels on the code embedding dataset. By evaluating the performance of the classification models, we aim to assess the correlation between code embedding metrics and the quality of code embeddings and gauge whether the established classification model is capable of discerning the quality of code embeddings based on their features, thereby determining whether the specific code PTM used to generate the embedding is suitable for use.

\section{Assess Recommendation Framework}\label{sec:assess}
Specifically, in the constructed code embedding dataset, we use the code embedding metrics values as data features 
and labels to establish five traditional machine learning models: LR, NB, RF, DT and RF. We divide the code embedding dataset with 8000 instances into the training set, validation set and test set in a ratio of 6:2:2 and then use the training set to train the five ML models. The performance of the classification model on the test set indicates the correlation between code embedding metrics and the quality of code embeddings and determines the recommended effects.

\begin{table}[!ht]
\small
  \caption{Performance of five machine learning models built with code embedding metrics.}
  \label{tab:result1}
\begin{tabular}{ccccc}
    \toprule
Models & Accuracy & F1-Score   & AUC   & MCC   \\
    \midrule
LR     & 0.790    & 0.688 & 0.787 & 0.542 \\
RF     & \textbf{0.906}    & \textbf{0.831} & \textbf{0.879} & \textbf{0.766} \\
NB     & 0.743    & 0.656 & 0.767 & 0.490 \\
DT     & 0.855    & 0.753 & 0.823 & 0.650 \\
KNN    & 0.875    & 0.781 & 0.839 & 0.695 \\
  \bottomrule
\end{tabular}
\vspace{-0.3cm}
\end{table}

\begin{figure}[!h]
  \centering
  \includegraphics[width=\linewidth]{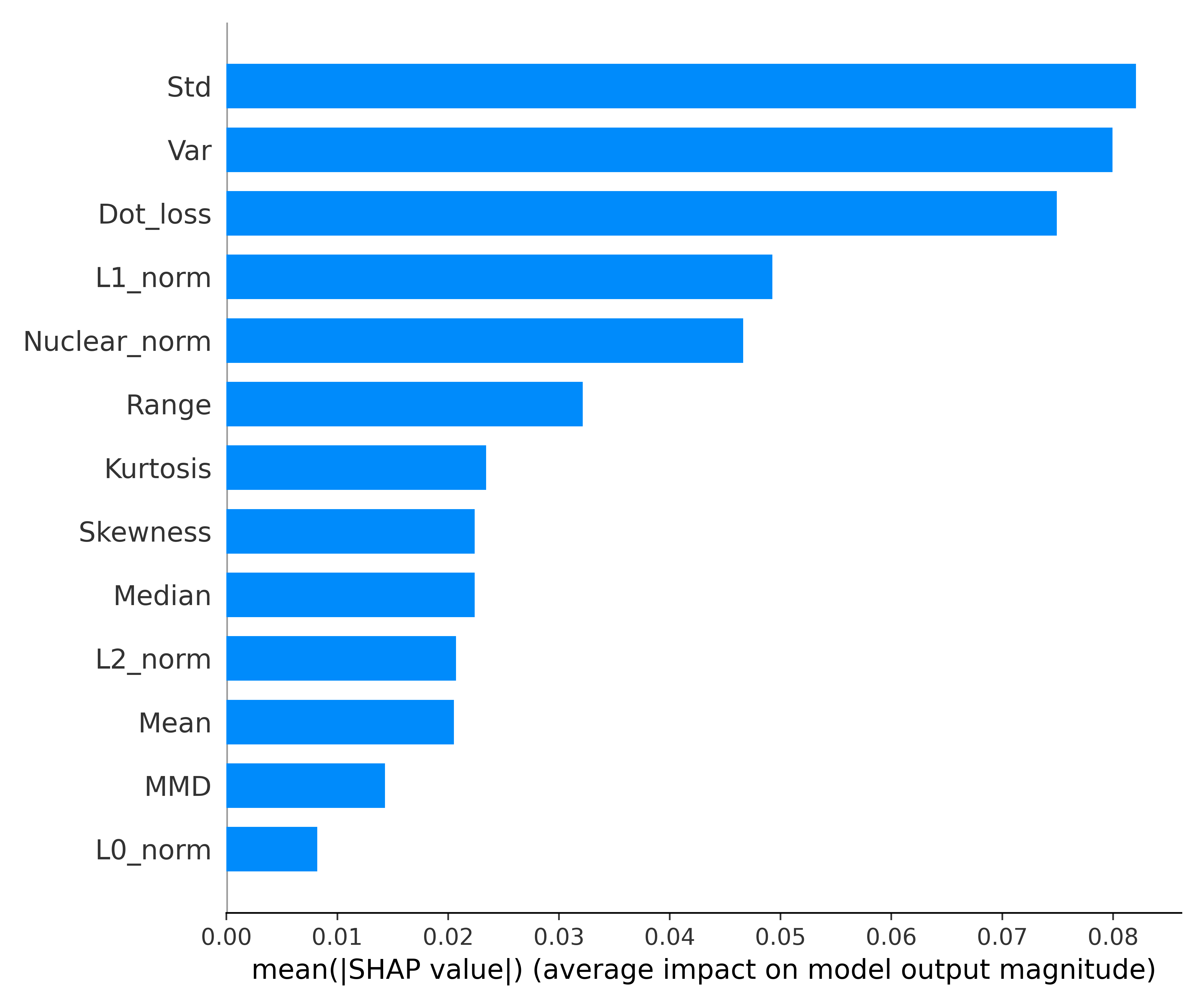}
  \caption{Feature importance rank of the RF model.}
\label{fig:discussion}
\vspace{-0.3cm}
\end{figure}

Table \ref{tab:result1} shows the performance of five machine learning models. As can be seen, \textbf{the defined code embedding metrics demonstrate a strong association with the quality of code embedding.} As depicted in Table \ref{tab:result1}, the performance of the five machine learning models, utilizing the code embedding metrics as features, on the test sets is notably high, with Accuracy, F1-Score, AUC, and MCC all exceeding 0.7, 0.65, 0.75, and 0.45 respectively. This indicates that the correlation between the code embedding metrics and the quality of code embedding is pervasive and can be effectively captured by various machine learning models. The established machine learning model can effectively discern whether the semantic richness contained within the code embedding ranks among the top three based on the code embedding metrics, thereby facilitating judgment regarding the suitability of the code PTM used for generating the code embedding. \textbf{Additionally, RF outperforms LR, NB, DT and KNN models in terms of Accuracy, F1-score, AUC, and MCC.} The superior performance of the RF model i.e. Accuracy of 91\%, F1-Score of 83\%, AUC of 88\% and MCC of 77\%, suggests its ability to establish a stronger correlation between the code embedding metrics and the quality of code embedding and to discern the quality of code embeddings based on their features, thereby determining whether the specific code PTM used to generate the embedding is suitable for use to achieve appropriate recommendation.

Next, we delve deeper into which code embedding metrics have made significant contributions to recommending code pre-trained models. To achieve this, we employ the feature importance method to quantify the contribution of each metric in the RF model. While some classifiers selected in this study may have their own model-specific methods for calculating feature importance \cite{PremrajH11, CalefatoLN19}, previous research has highlighted inconsistencies in determining the most important features using such model-specific approaches. Model-independent methods like SHapley Additive exPlanations (SHAP) \cite{LundbergL17} and Permutation \cite{Tantithamthavorn20} have demonstrated high consistency in calculating feature importance \cite{rajbahadur2021impact}. Given this, we opt for the model-independent feature importance method i.e. SHAP to calculate feature importance. SHAP has been theoretically proven to provide the best feature importance ranking and is gaining traction in the software engineering community \cite{Jiarpakdee19, iarpakdeeTT20}. 

Figure \ref{fig:discussion} shows the feature importance rank of the RF model in RQ3. As shown in Figure \ref{fig:discussion}, each metric designed contributes to the prediction process of the RF model. First, the metric with the largest contribution is the standard deviation. In addition, the importance of variance is very close to the standard deviation. This result is obvious from the mathematical principle. In fact, the two most important metrics are defined from the perspective of statistical characteristics, which corresponds to the first two characteristics of the code embedding differences observed in RQ2. The three related metrics, namely, numerical range, skewness and kurtosis, are at the sixth, seventh and eighth levels respectively. Secondly, the dot product metric emerges as the third most influential metric. Defined from the perspective of distribution, its purpose is to measure distribution differences. MMD, serving a similar purpose, holds lower importance, indicating that the data distribution calculated based on vector granularity levels in the code embedding better reflects differences between different code embeddings. Finally, L1 norm and Nuclear norm are the fourth and fifth most important features, which are defined from the perspective of norm and also contribute greatly to the prediction process of the model.

\section{Apply in Practice}\label{sec:apply}
As shown in Section \ref{sec:assess}, the RF model shows the best recommendation effect. So next we are ready to apply this model to real scenarios. However, since it is difficult to obtain high-quality open-source code vulnerability datasets, we rebuilt the recommended framework on three of the four datasets selected in this paper and applied it to the remaining one dataset. In more detail, for the process of dividing the dataset and building the model in Section \ref{sec:assess}, we will use the code embedding dataset of 6000 instances built based on the three datasets of Devign, CWE119 and CWE399 as the training set. The constructed RF model is used as the recommendation framework to make recommendations for a new unseen reveal dataset. In order to ensure the authenticity of practical applications, we also sampled the Reveal dataset 200 times to obtain 200 new datasets for practical testing. We obtain the top three code PTMs with the best performance on these 200 new datasets using a process similar to Section \ref{sec:preliminary_study}, and then we compare whether the code PTMs recommended by the recommendation framework for these new datasets are consistent with reality.

\textbf{Experimental results show that our recommendation framework achieves an accuracy of 78\% for the code pre-trained model recommended for these 200 unseen new datasets.} The pre-trained models recommended by the recommendation framework for the 156 new datasets were consistent with the pre-trained models that actually achieved the best performance. This means that our recommendation framework can successfully recommend code pre-trained models for new datasets in the same task domain, and the pre-trained model can obtain higher-quality code embeddings, resulting in better performance.

\section{Guidelines for Using Our Framework}\label{sec:discussion}

\begin{figure}[!ht]
\vspace{-0.3cm}
  \centering
  \includegraphics[width=\linewidth]{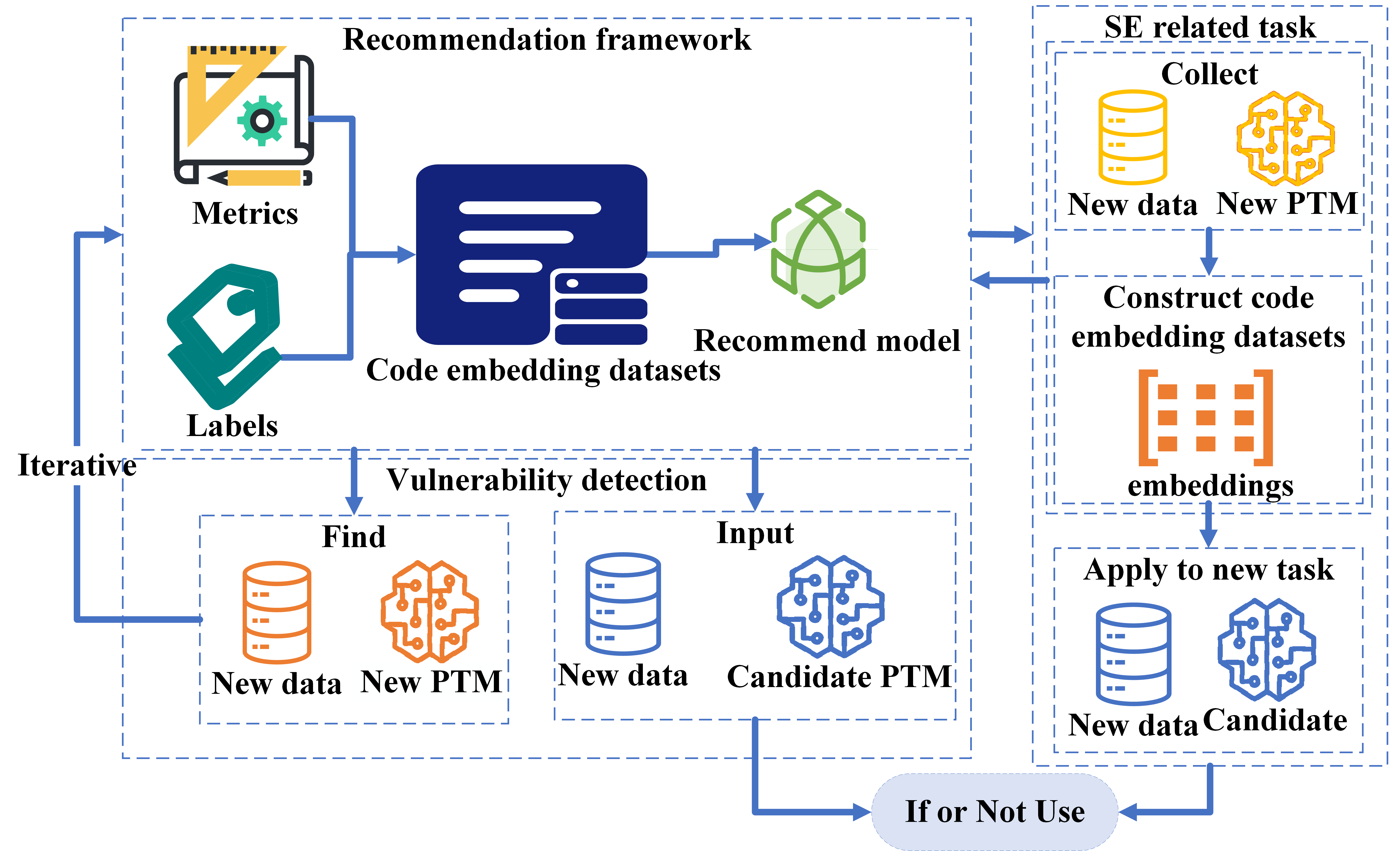}
  \caption{Guidelines for using our framework.}
\label{fig:conclude}
\vspace{-0.55cm}
\end{figure}

We explain our proposed framework in detail in Section \ref{sec:proposed_method} and evaluate the effectiveness of the proposed framework in Section \ref{sec:assess} and \ref{sec:apply}. Additionally, this section provides practical guidance on how to use, iterate, and build our recommendation framework within the vulnerability detection task domain and in other code-related task domains. Figure \ref{fig:conclude} shows the overall workflow of our framework.

1) We recommend using our framework to quantify and evaluate the code embedding representations produced by different pre-trained models, especially if you want to observe the difference between the code datasets at hand after vectorization.

2) Within the vulnerability detection task domain, when researchers and practitioners hesitate among multiple code pre-trained models. We propose to directly use our recommendation framework to recommend suitable code pre-trained models for vulnerability detection datasets to generate high-quality code embeddings.

3) Within the vulnerability detection task domain, when researchers and practitioners discover the best code pre-trained models on certain datasets. We propose to add the corresponding code embedding features and labels extracted by the proposed code embedding metric to the code embedding dataset of this article and retrain the recommendation model to extend the applicable scope.

4) For other SE-related tasks, we recommend using the framework construction process proposed in this paper to build a recommendation framework in the corresponding task field to select an appropriate code pre-trained model.

\section{Threats to Validity}\label{sec:threats_to_validity}
Although the experiments in this study demonstrate the effectiveness of our method, the broader validity of these findings may still be subject to certain limitations.

Most of the threats to the validity come from the external validity. The conclusion may be mostly valid for code PTMs and the specific SE  tasks we choose. We tried to minimize this threat by using four code PTMs covering all three different architectures for generating ten different code embeddings. Moreover, it is worth noting that it is necessary to expand experiments on the most advanced and more code PTMs and SE downstream tasks.

The biggest threats to the internal validity are the datasets and performance evaluation metrics. We counteract this threat by utilizing publicly available datasets widely used by previous researchers and evaluating performance with Accuracy, F1-score, AUC and MCC four metrics.

A possible threat to construct validity regards the methodology used to evaluate the code embeddings. Since we primarily focus on code embeddings, when we obtain the embeddings we only train and fine-tune the same fully connected layer for classification across different embeddings and get the single AUC performance value to gauge the quality of code embeddings, which is the same as previous researchers \cite{sharma-attention-bert, wang2024empirical, dou2023towards}. In addition, we use the code embedding metrics as data features to build five traditional machine learning models from different families on the newly constructed code embedding dataset and analyze the effectiveness of the metrics by observing the performance of these different classifiers, which is also an obvious common practice. However, more comprehensive and reasonable methods should also be explored in the future for the above two aspects.

\section{Conclusion}\label{sec:conclusions}
To ensure software security, software vulnerability detection is a key research area in the software engineering field all the time. 
This paper first systematically investigates the impact of code embeddings generated by different code PTMs on vulnerability detection tasks. Then further define a set of code embedding metrics to quantify the differences and characteristics between these code embeddings. Finally, proposes a recommendation framework based on the metrics to guide researchers in selecting appropriate code PTMs to generate high-quality code embeddings for better performance. 
Our case study shows that the defined embedding metrics demonstrate a strong association with the embedding quality and the recommendation framework based on the metrics can achieve an Accuracy of 91\%, F1-Score of 83\%, AUC of 88\% and MCC of 77\% to recommend appropriate code PTMs on the test sets. When applied in practice, the PTMs recommended by the framework for the 156/200 new unseen datasets were consistent with the PTMs that actually achieved the best performance. Our research results provide clues for future SE researchers and practitioners to select appropriate code PTMs from the PTM repository to generate high-quality code embeddings for code-related tasks.


\bibliographystyle{plain}
\bibliography{sample-sigconf-biblatex} 

\end{document}